\documentclass[doublecol,linenumbers]{epl2}
\usepackage{graphicx,grffile}
\usepackage{dcolumn}
\usepackage{bm}
\usepackage{subfig}
\usepackage{caption}
\usepackage{xcolor}
\usepackage{float}
\usepackage{amsmath,amssymb}
\nolinenumbers
\begin{document}
\title{Fragility of Chimera States under Dynamic Random Links}
\title{Dynamic Random Links destroy Chimera States}
\title{Chimera States are Fragile under Random Links}
\author{Sudeshna Sinha}
\institute{Indian Institute of Science Education and Research Mohali,\\ Knowledge 
City, SAS Nagar, Sector 81, Manauli PO 140 306, Punjab, India \footnote{sudeshna
@iisermohali.ac.in}}
\abstract{
We study the dynamics of coupled systems, ranging from maps supporting chaotic attractors to nonlinear differential equations yielding limit cycles, under different coupling classes, connectivity ranges and initial states. Our focus is the robustness of chimera states in the presence of a few time-varying random links, and we demonstrate that chimera states are often destroyed, yielding either spatiotemporal fixed points or spatiotemporal chaos, in the presence of even a single dynamically changing random connection. We also study the global impact of random links by exploring the Basin Stability of the chimera state, and we find that the basin size of the chimera state rapidly falls to zero under increasing fraction of random links. This indicates the extreme fragility of chimera patterns under minimal spatial randomness in many systems, significantly impacting the potential observability of chimera states in naturally occurring scenarios.
}

\maketitle

Coupled dynamical systems have provided a wide class of simple models that have significantly captured the essential features of large interactive complex systems \cite{cml,cml2,chate,chate2}. Such spatially distributed systems have provided frameworks for understanding and characterizing spatiotemporal patterns emerging in problems ranging from multimode lasers and coupled Josephson Junctions, to microfluidic arrays and evolutionary biology \cite{example1,example2,example3,example4}.

A particular spatiotemporal pattern, the chimera state, has caught widespread research attention in recent years, in fields ranging from physics and chemistry, to biology and engineering [9-29]. One of simplest examples of a chimera state \cite{chimera} is a ring of coupled identical phase oscillators which spontaneously breaks the underlying symmetry and splits into synchronized and desynchronized groups. While such fascinating patterns had been observed in many systems in the past \cite{cml,cml2,chate,chate2,liqcrys}, they have been dubbed a ``chimera'' in recent times \cite{chimera1}. In particular chimera-like phenomena have been seen in numerical simulations, as well as some experimental realizations, such as Josephson junction arrays \cite{chimera16}, star networks \cite{star}, electrochemical systems \cite{chimera17,chimera18}, uni-hemispheric sleep \cite{chimera18}, electronic circuits\cite{star,chimera20}, optical analogs of coupled map lattices \cite{chimera9}, mechanical metronomes \cite{chimera19} and Belousov-Zhabotinsky chemical oscillator systems \cite{chimera8}. A very pertinent issue for the observability of chimeras is the robustness of these patterns, and significant understanding of chimeras under varying connection topologies and inhomogenieties has been obtained in Refs.~\cite{scholl1,scholl2}. In this work we will extend this understanding, by focusing on the persistence of chimera patterns in coupled nonlinear systems under time-varying random links. The surprising leading result here is the following: when a few of the regular connections in these systems are dynamically randomized, the chimera states are destroyed and the symmetry-breaking spatial patterns are eliminated.  Namely, the {\em chimera states are very fragile under dynamic random links.}  Since in many systems of physical, technological and biological significance a certain degree of randomness in spatial links is closer to physical reality \cite{Watts}, our finding that random links kill chimeras is significant and suggests a generic underlying mechanism due to which complex systems may not exhibit chimera states.

As a sufficiently general test-bed we will consider a range of generic coupled systems, comprised of nonlinear local dynamics and a coupling term modelling the interaction, and we will demonstrate our central result in different classes of such systems. First we will consider coupled circle maps and coupled bistable maps under diffusive coupling, varying in range from local intercations to connections spanning a large set of neighbours. We will also go on to explore the robustness of chimeras in coupled oscillator systems, considering another class of coupling, namely conjugate coupling through dissimilar variables. We will consider examples where the chimera state arises from generic random initial states in a ring, as well as situations where special initial states give rise to chimeras. The salient question here is the resilience of the chimera state under dilute time-varying random links in these systems.\\

\medskip

\noindent
{\bf Coupled Map Lattices:}\\

Our first representative example is in the general class of {\em Coupled Map Lattices} (CML) \cite{cml,cml2}, where the local dynamics at the sites is described by a {\em circle map}.  The maps are coupled to two nearest neighbours through {\em diffusive coupling}, and such a system has been used to model the behavior of coupled oscillators, such  as Josephson Junction arrays. The dynamical equations for such coupled maps are:
\begin{equation}
  x_{t+1} (i) = (1 - \epsilon ) f(x_t (i)) + \frac{\epsilon}{2} \{ f(x_t (i+1)) + f(x_t (i-1)) \}
\label{circle_cml}
\end{equation}
where the on-site dynamics is given by $f(x) = x + \Omega - \frac{K}{2 \pi} \sin (2 \pi x)$. The dynamics above is defined modulo $1$, i.e. it maps a circle onto itself. The two significant parameters in the system is $\Omega$ which can be interpreted as an externally applied frequency, and $K$ which reflects the strength of the nonlinearity \cite{circle}.

Starting with connections given by a ring topology, we consider increasingly random networks formed dynamically as follows: a fraction $p$ of the regular links are replaced by random connections, i.e. ``rewired randomly''. This implies that at any instant of time a fraction $p$ of random links co-exist along-side regular links. Such networks have been seen to have widespread relevance to a range of natural and engineered phenomena \cite{Watts}. Note that our coupling occurs on a degree preserving directed network. Further, we consider the random links to be time-varying here. So the underlying web of connections changes over time, with a (typically small) number of links dynamically rewired randomly from time to time \cite{dynamic_links1,dynamic_links2,dynamic1,dynamic2}. For maps, the links switch at every iteration $t$. For networks of coupled nonlinear systems described by differential equations, which we will consider later, the rewiring takes place at very short time intervals vis-a-vis the intrinsic time-period of the constituent oscillators. The results presented here are robust to a wide range of rewiring frequencies. Such dynamic connections are expected to be widely prevelant in complex systems, for instance in scenarios where links change from time to time in response and adaptations to external environmental factors or internal influences \cite{dynamic1,dynamic2}.

\begin{figure}[h]
  \centering
  \includegraphics[width=0.55\linewidth,height=0.9\linewidth,angle=270]{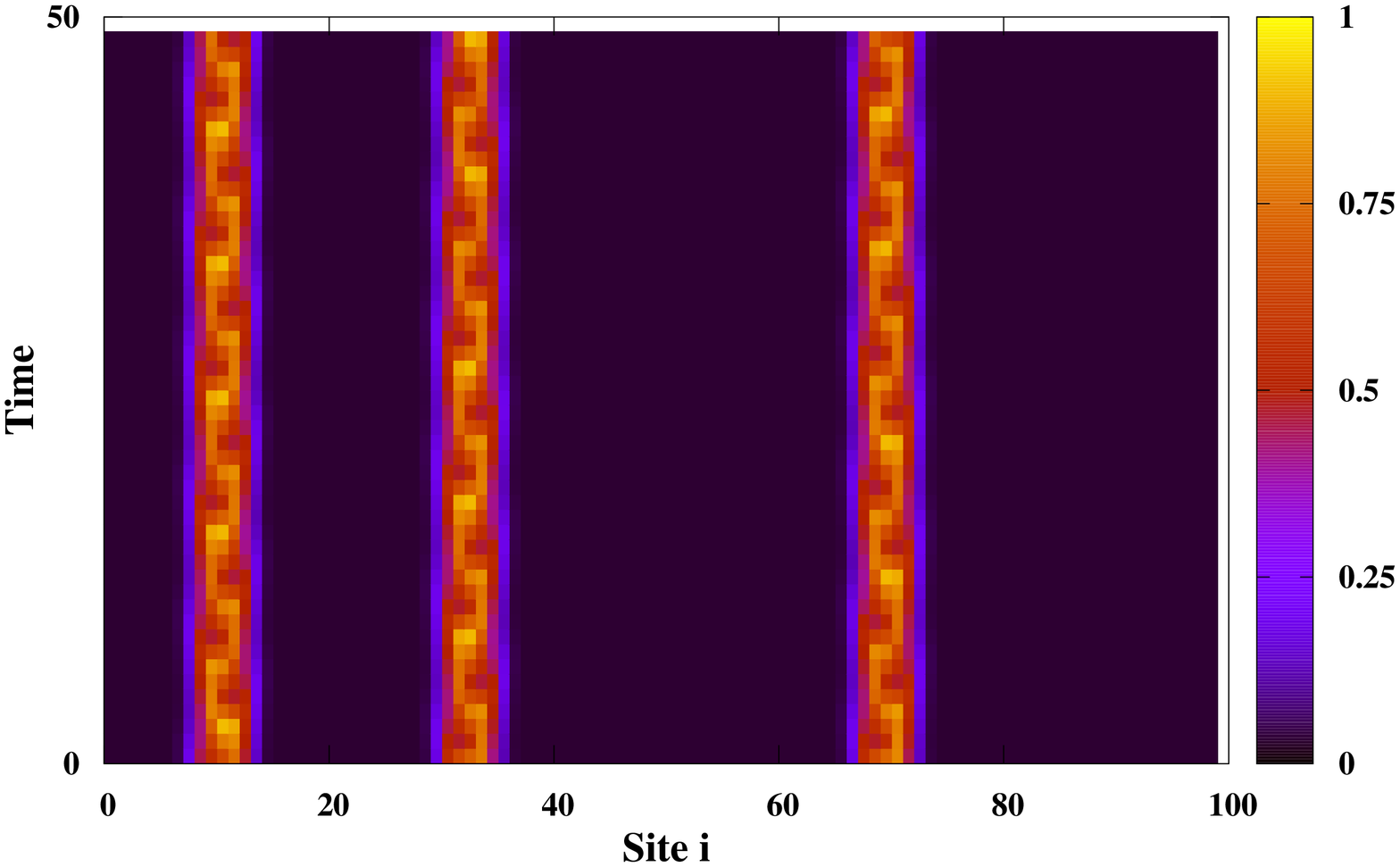}
  \includegraphics[width=0.55\linewidth,height=0.9\linewidth,angle=270]{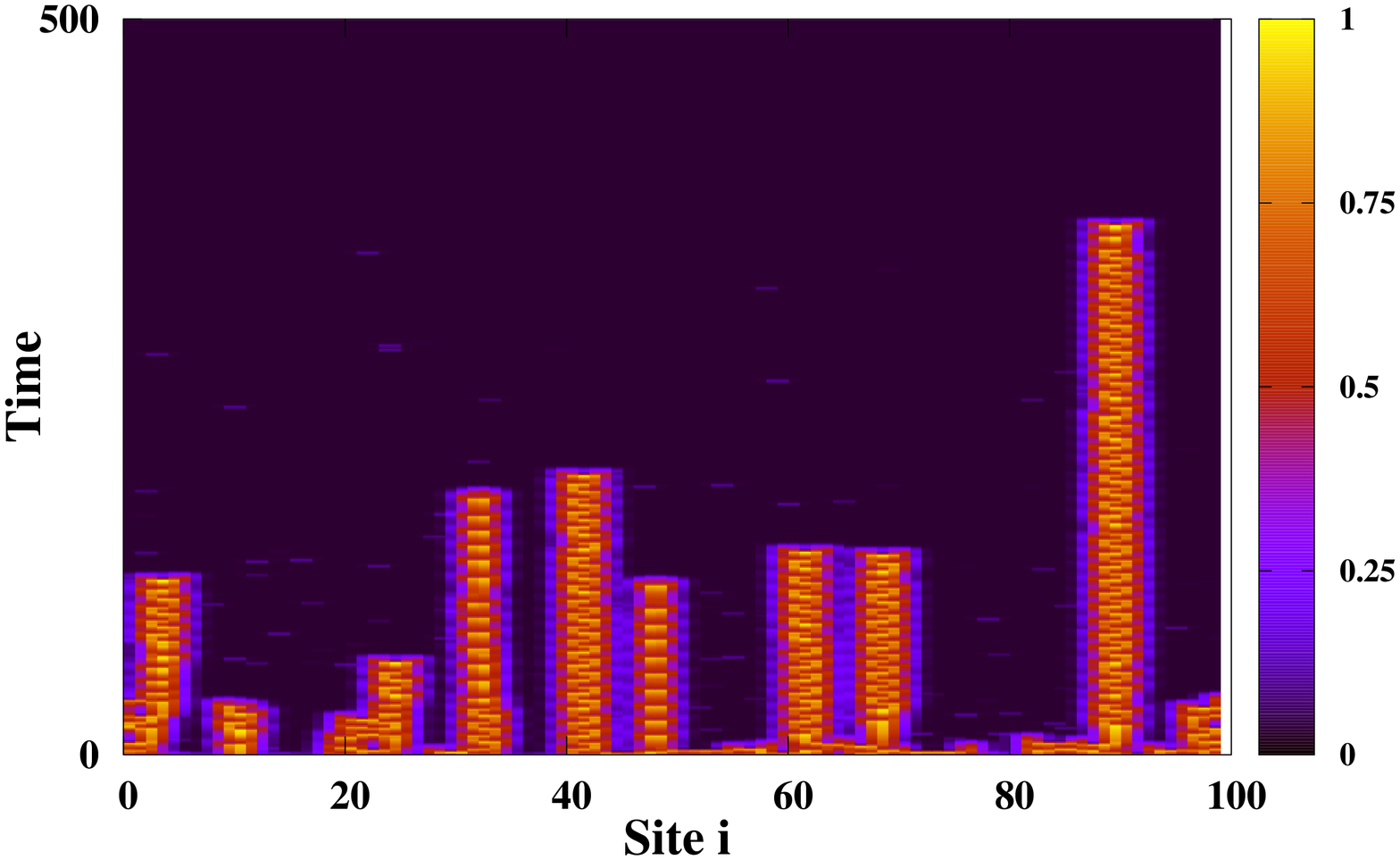}
  \caption{Evolution of coupled circle maps (with $K=1$, $\Omega = 0.031$, $\epsilon = 0.42$ in Eqn.~\ref{circle_cml} and system size $N=100$). The top panel shows a regular ring ($p=0$), after transience of $450$ iterations. The lower panel shows the case of $p=0.01$ where, on an average, there is a {\em single} random link in the entire ring. Both cases evolve from the {\em same random initial state}, with the initial state drawn from an uniform random distribution in the range $[0:1]$. Notice that the chimera-like pattern present in a ring (top panel) is destroyed by the single dynamic random link to yield the sptaiotemporal fixed point in (lower panel).\\}
\label{circle_p0.01}
\label{circle1}
\end{figure}

The evolution of spatiotemporal patterns for this system of coupled circle maps is displayed in Fig.~\ref{circle1}, for the illustrative case of $K=1$, $\Omega = 0.031$, $\epsilon = 0.42$ in Eqn.~\ref{circle_cml}. The first case is a regular ring of coupled maps (see top panel of Fig.~\ref{circle1}), while the second case has a {\em single} link rewired randomly (see lower panel of Fig.~\ref{circle1}). Both cases evolve from the {\em same random initial state}, i.e. the set of $x(i)$, $i=1, \dots N$, at time $t=0$ is identical for both cases. However, very clearly, the dynamical outcome in the two cases is drastically different. So the presence of a {\em single} dynamically changing random link in the ring destroys the chimera-like pattern observed in the ring, instead yielding a homogeneous steady state (i.e. a spatiotemporal fixed point where $x (i)= x^{\star}$, for all $i$).

\begin{figure}[htb]
  \centering
  \includegraphics[width=0.45\linewidth,height=0.75\linewidth,angle=270]{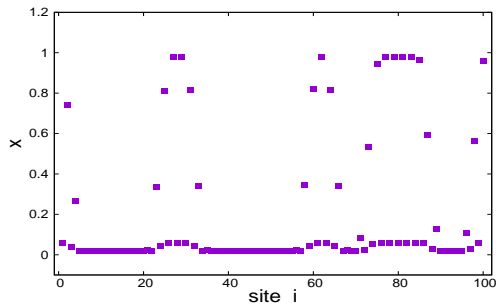}
  \caption{Chimera-like spatial profile $x(i)$ of coupled circle maps ($i =1, \dots 100$), after 500 transient time steps, for a regular ring (i.e $p=0$). Here $K=1$, $\Omega = 0.019$, $\epsilon = 0.9616$ in Eqn.~\ref{circle_cml}.\\}
  \label{circle_spatial_p0}
\end{figure}

We present another representative example of the destruction of a chimera state by a single random link. Here the parameters in Eqn.~\ref{circle_cml} are $K=1$, $\Omega = 0.019$, $\epsilon = 0.9616$. Again it is clearly evident that the presence of a {\em single} random link in the ring destroys the chimera-like pattern observed in the ring (see Fig.~\ref{circle_spatial_p0}), yielding a spatiotemporal fixed point instead (see Fig.~\ref{circle_p0.01}).

\begin{figure}[htb]
    \centering
    \includegraphics[width=0.55\linewidth,height=0.9\linewidth,angle=270]{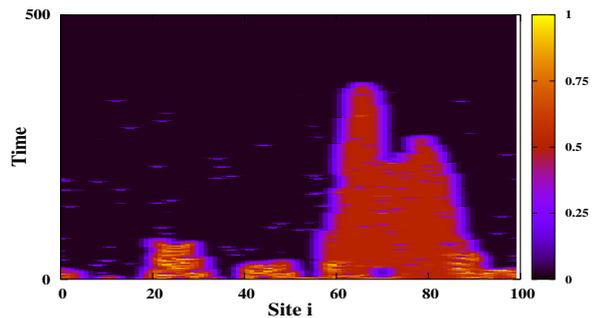}
\caption{Time evolution of $x_t (i)$ of  coupled circle maps, for fraction of random links $p=0.01$ , i.e. when there is a {\em single} random link in the entire ring. Here $K=1$, $\Omega = 0.019$, $\epsilon = 0.9616$ in Eqn.~\ref{circle_cml} and system size $N = 100$. The initial conditions here are the same as in Fig.~\ref{circle_spatial_p0}. However, in contrast to the chimera-like pattern in  Fig.~\ref{circle_spatial_p0}, we obtain a sptaiotemporal fixed point here.\\}
\label{circle_p0.01}
 \end{figure}

\begin{figure}[htb]
    \centering
  \includegraphics[width=0.45\linewidth,height=0.75\linewidth,angle=270]{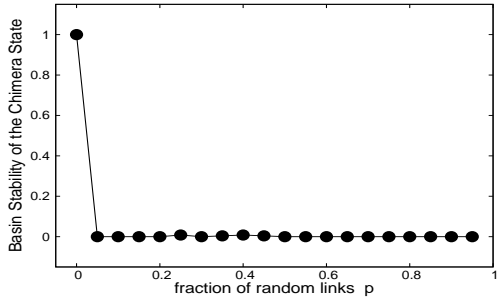}
  \caption{Basin Stability of the chimera state in coupled circle maps, where Basin Stability is estimated from the fraction of initial states that go to the chimera state. Here $K=1$, $\Omega = 0.019$, $\epsilon = 0.9616$ in Eqn.~\ref{circle_cml}, and Basin Stability is obtained by sampling $250$ random initial conditions. State variables are considered synchronized if they are similar within an accuracy of $10^{-6}$, after transience of $10^3$. Identical results are obtained for the parameter set $K=1$, $\Omega = 0.031$, $\epsilon = 0.42$.\\}
\label{bs}
 \end{figure}

In order to quantify the global impact of random links we explore the Basin Stability \cite{bs1,bs2} of the chimera state, which reflects the probability of a generic initial state evolving to a chimera state. The Basin Stability is estimated by sampling a large set of initial states and assessing the number of initial states that yield chimeras asymptotically. If this fraction is close to $1$, we can deduce that a generic randomly chosen initial state will evolve to a chimera state with probability close to one, and if it is close to zero, it implies that almost no random initial state will yield a chimera. Fig.~\ref{bs} shows this fraction as a function of the fraction of random links $p$ in the ring. It is clearly seen that the basin size of the chimera state sharply decreases to zero as the fraction of random links becomes non-zero, i.e. we obtain a sharp transition to generic non-chimera states as $p \rightarrow 0$. Identical Basin Stability results are obtained for the two parameter sets investigated. So in this CML a generic randomly chosen initial state will almost certainly yield a chimera state for a regular ring, but {\em one} random link will almost certainly destroy the chimera pattern and yield a non-chimera state.

Interestingly, the non-chimera states are spatiotemporal fixed points for $p < 0.5$, while for higher $p$ the chimeras are destroyed yielding weak spatiotemporal chaos. So this system of coupled circle maps yields three kinds of dynamical states: (i) ``chimera-like'' states for the special case of the regular ring, (ii) spatiotemporal fixed points for a small number of random links, and (iii) spatiotemporal chaos when the number of random links is predominant. This suggests that the chimera state is very delicate and occurs only in the limit of a completely regular ring.

An important aspect to note here is the following: linear stability analysis cannot be employed to understand this phenomena. The spatiotemporal fixed point is in fact linearly stable for small $p$, and initial conditions very close to the fixed point indeed evolve quickly to a homogeneous steady state. However, in terms of global stability the situation is interesting and non-trivial under varying degrees of randomness. For the regular ring (i.e. $p=0$) the basin of attraction of the homogeneous fixed point is localized close to the fixed point, and a very large set of initial conditions away from this narrow band in state space go to chimera states. So when the state space is randomly sampled this set dominates and the Basin Stability of the chimera state tends to one. However counter-intuitively, when even a single link is randomized, the basin of the spatiotemporal fixed point grows explosively to a near global attractor (within the limits of numerical sampling and accuracy), and the basin of chimera states shrinks drastically. So our observed phenomena is crucially dependent on global stability considerations, which is typically analytically intractable.

In order to check the generality of these observations, we now explore another class of coupled map lattices where the {\em coupling range is not restricted to nearest neighbours}. Rather the coupling extends to $k$ neighbours on both sides, where $k > 1$ \cite{pw}.

The local dynamics is also chosen to be in a different class in order to explore a wider set of systems, and ascertain the generality of our results. So now we go on to consider coupled systems that are locally bistable, with stable fixed points co-existing with {\em chaotic attractors}. Our particular example is a ring of coupled piecewise-linear maps given as follows:

\begin{equation}
x_{t+1} (i) = f(x_t (i)) + \frac{\epsilon}{2k} \sum_{j=i-k}^{i+k} \{ f(x_t (j)) - f(x_t (i)) \}
\label{pw_cml}
\end{equation}
where the on-site dynamics is given by
\begin{eqnarray*}
  f(x) = \begin{cases} p_1 \ x \ + \ (p_1/l - 1)  \ \ \ \ \ \  \ \ \ \ \ x \in [-1, -1/l),\\ 
     l \ x  \ \ \ \ \ \ \ \ \ \ \ \ \ \ \  \ \ \ \ \ \ \ \ \ \ \  \ \ \ \ \  x \in [-1/l, 1/l),\\
      p_2 \ x \ - \ (p_2/l -1)   \ \ \ \ \ \  \  \ \ \ \   x \in [1/l, 1]
\end{cases}
\end{eqnarray*}  
The parameters $p_1$, $l$ and $p_2$ determine the slopes of the linear segments in different ranges of the state variable $x$. We
choose the parameters such that a stable fixed point co-exists with a chaotic attractor \cite{pw}. Specifically, we consider $p_1 = -0.5$, $p_2 = -2.4$, and $l = 1.5$. Here the local dynamics supports a steady state at $-8/9$ whose basin of attraction is the interval $[-1, 0)$. It also supports a chaotic attractor with span $[0.2, 1]$, having a  basin of attraction in the interval $(0, 1]$.

\begin{figure}[h]
  \centering
    \includegraphics[width=0.4\linewidth,height=0.75\linewidth,angle=270]{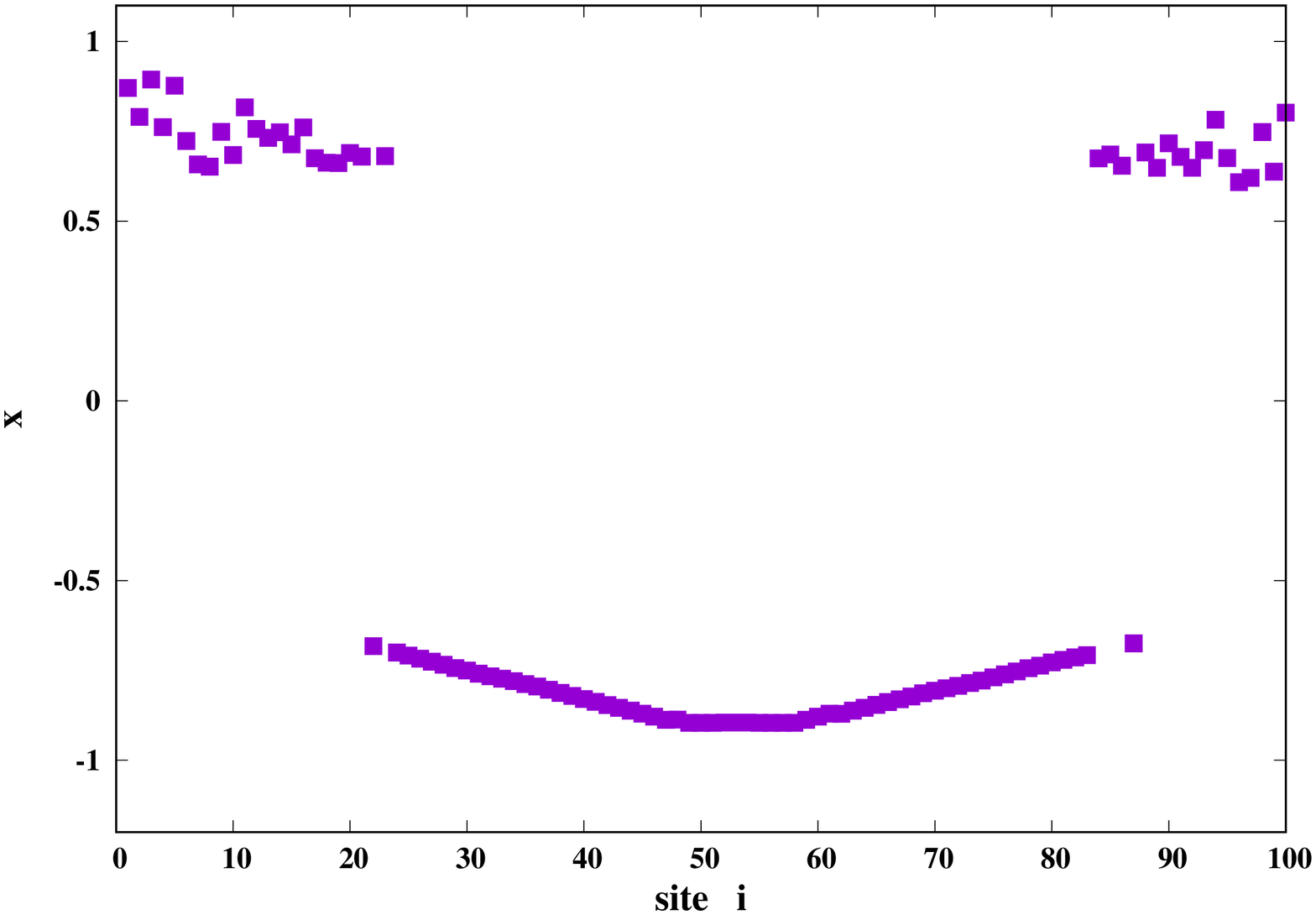} 
  \includegraphics[width=0.45\linewidth,height=0.8\linewidth,angle=270]{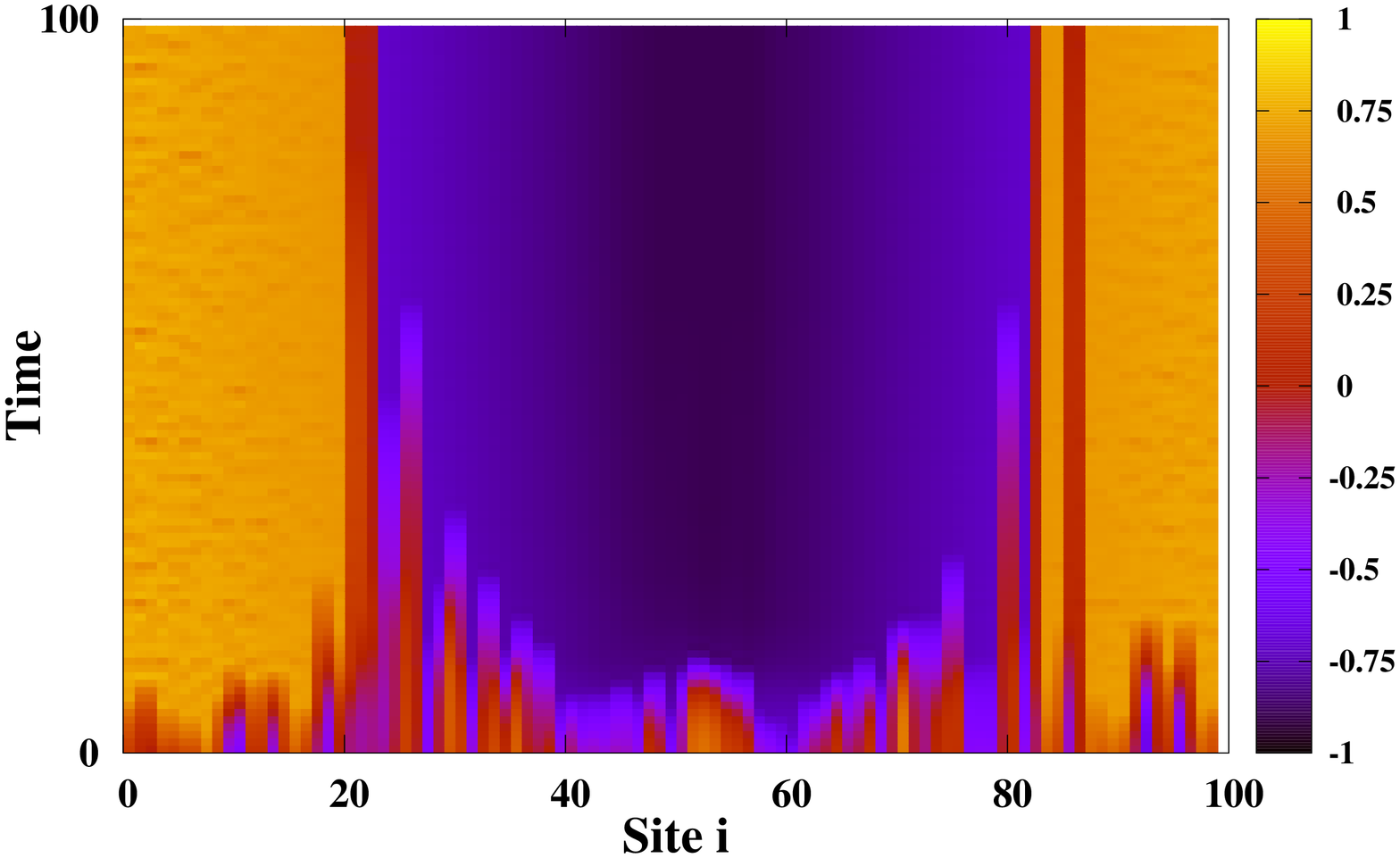}  
  \includegraphics[width=0.45\linewidth,height=0.8\linewidth,angle=270]{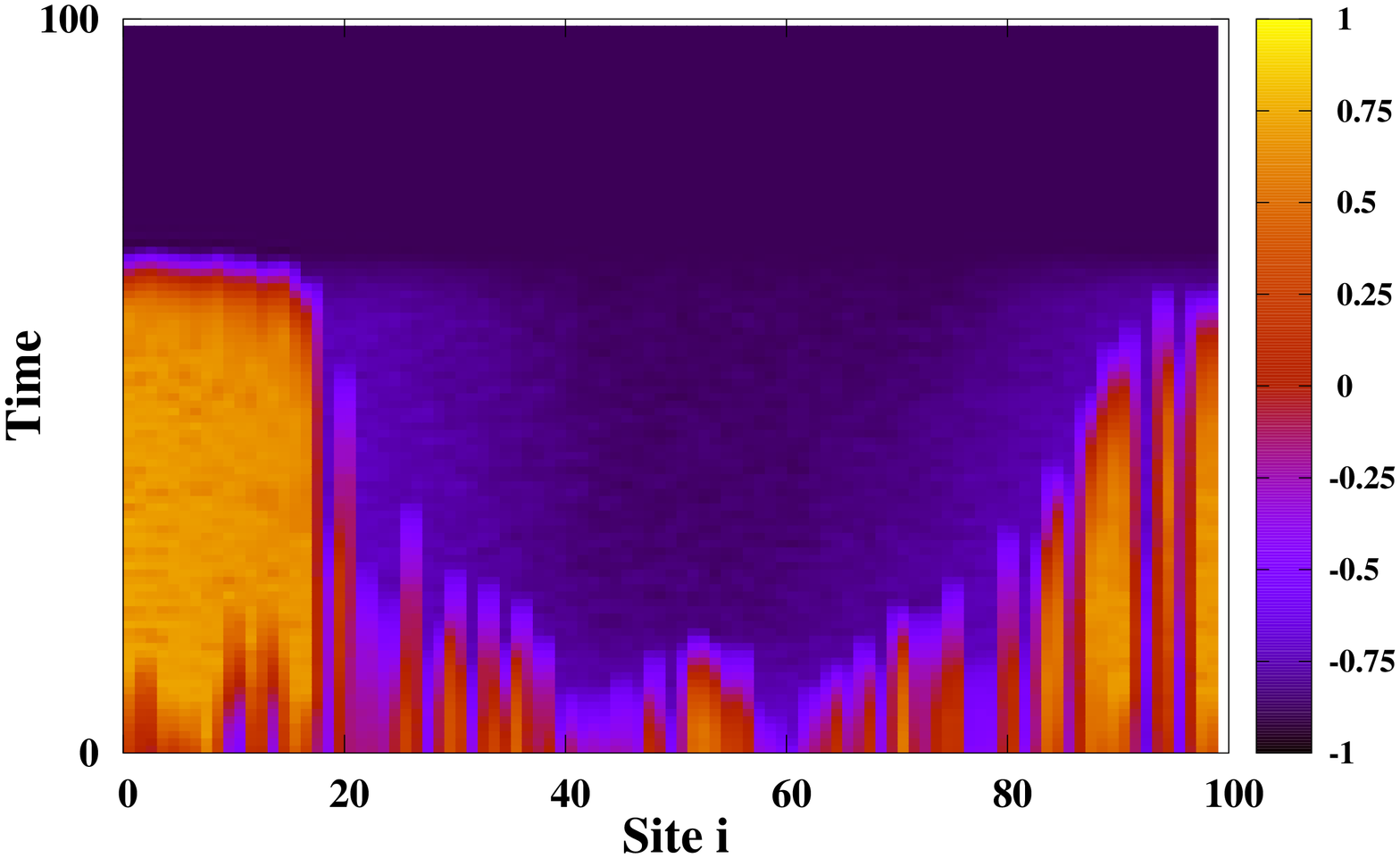}
  \caption{Spatial profile $x (i)$ of coupled piecewise-linear bistable maps ($i =1, \dots 100$), after 500 transient time steps (top panel); time evolution of $x_t (i)$ for the case of a regular ring (middle panel) and $p=0.3$ (lower panel). The initial states are identical in all panels. Parameters in Eqn.~\ref{pw_cml} are $p_1 = -0.5$, $p_2 = -2.4$, $l = 1.5$, $k=25$ and $\epsilon=0.3$. Notice that the emergent chimera-like pattern in (top panel) is destroyed under some fraction of random links in the ring to yield  a spatiotemporal fixed point in (lower panel).\\}
\label{pw_eps_0.3}
\end{figure}

The evolution of spatiotemporal patterns for this system of coupled piecewise-linear maps is displayed in Fig.~\ref{pw_eps_0.3}, for two illustrative cases. The first case is a regular ring of coupled maps (see top and middle panels of Fig.~\ref{pw_eps_0.3}), while the second case has a fraction of links rewired randomly (see lower panel of Fig.~\ref{pw_eps_0.3}). Both cases evolve from the {\em same random initial state}, i.e. the set of $x(i)$, $i=1, \dots N$, at time $t=0$ is identical for both cases. It is clearly evident that the dynamical outcome in the two cases is again drastically different and the presence of random links again destroys the chimera-like pattern observed in the ring, yielding a spatiotemporal fixed point instead.

\begin{figure}[h]
\centering
  \includegraphics[width=0.45\linewidth,height=0.85\linewidth,angle=270]{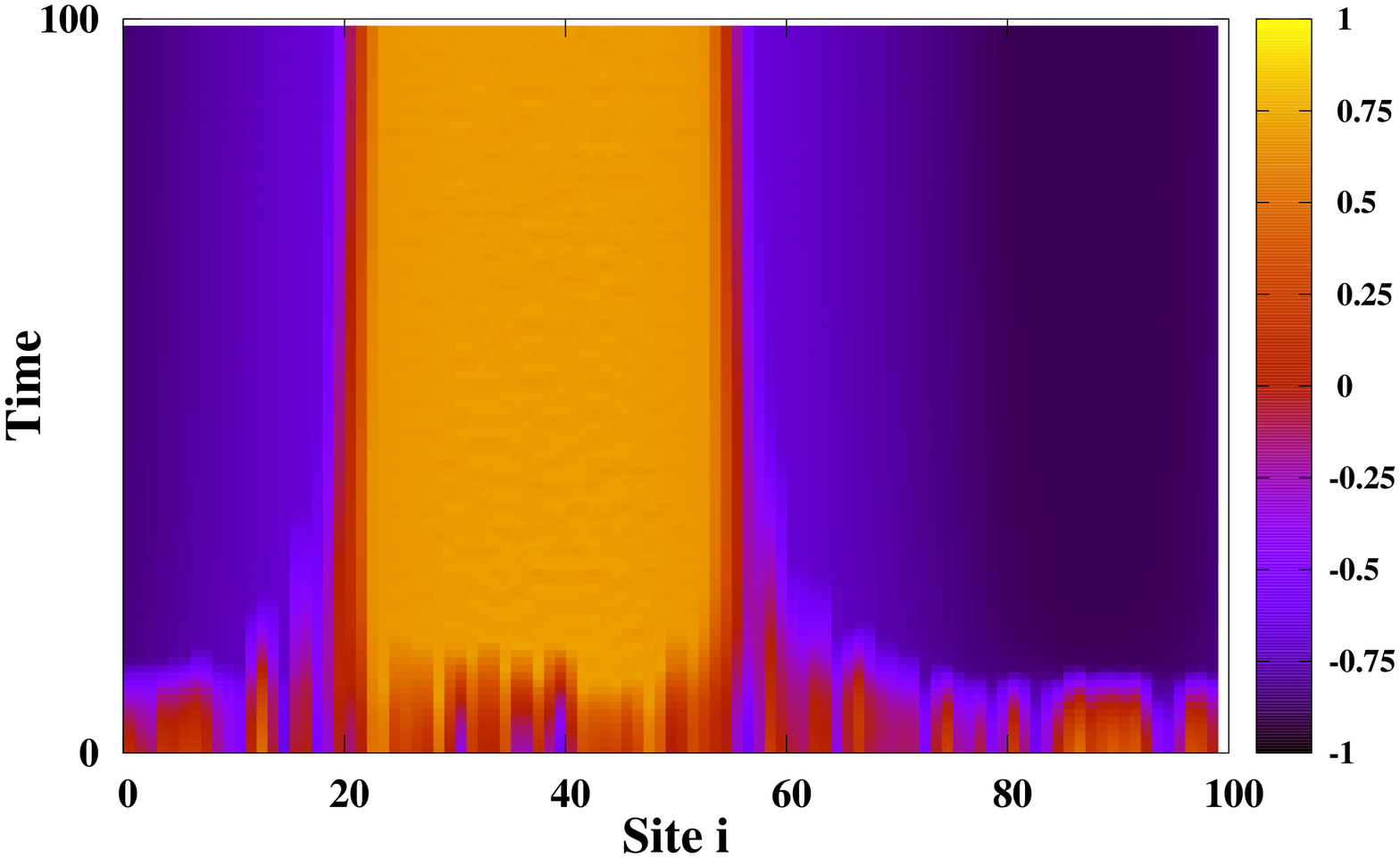} 
    \includegraphics[width=0.45\linewidth,height=0.85\linewidth,angle=270]{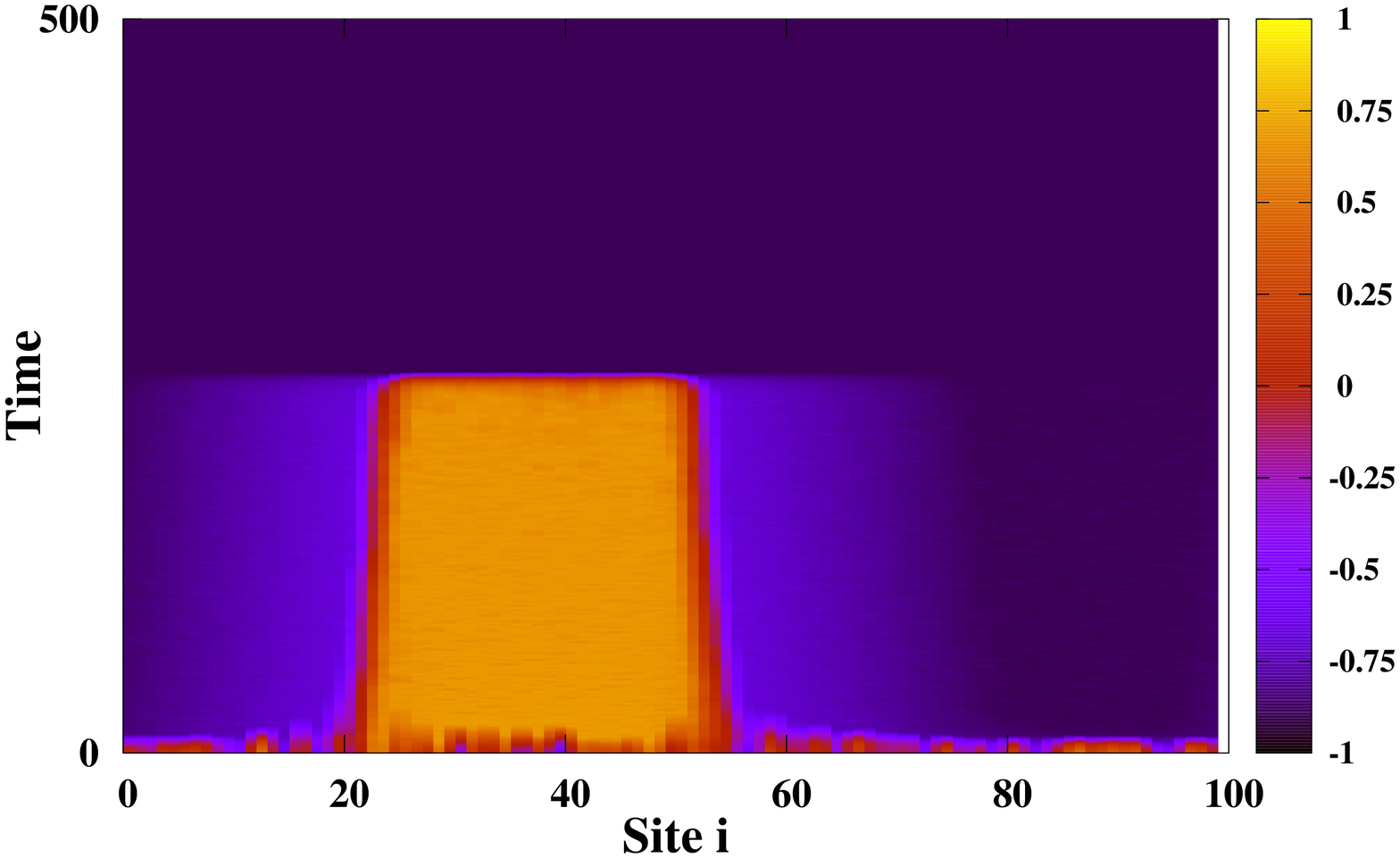} 
    \includegraphics[width=0.45\linewidth,height=0.85\linewidth,angle=270]{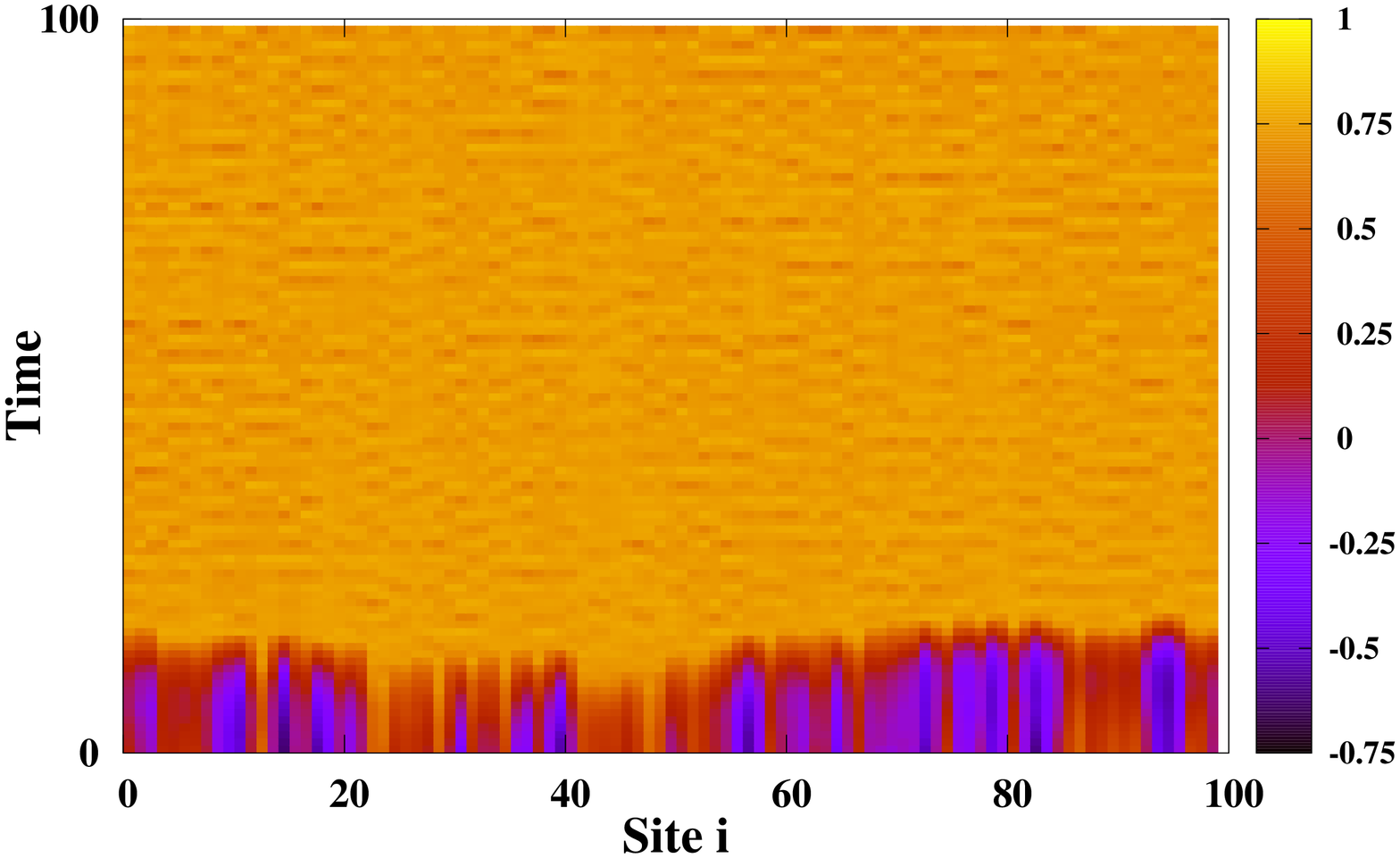}
\caption{Evolution of coupled piecewise-linear bistable maps for fraction of random links (top) $p=0$, (middle) $p=0.05$, (lower) $p=0.3$. Here system size $N=100$ and the parameters in Eqn.~\ref{pw_cml} are $p_1 = -0.5$, $p_2 = -2.4$, $l = 1.5$, $k=25$ and $\epsilon=0.35$. The initial state is the same in all panels.\\}
\label{pw_eps_0.35}
\end{figure}

\begin{figure}[h]
\centering
  \includegraphics[width=0.45\linewidth,height=0.75\linewidth,angle=270]{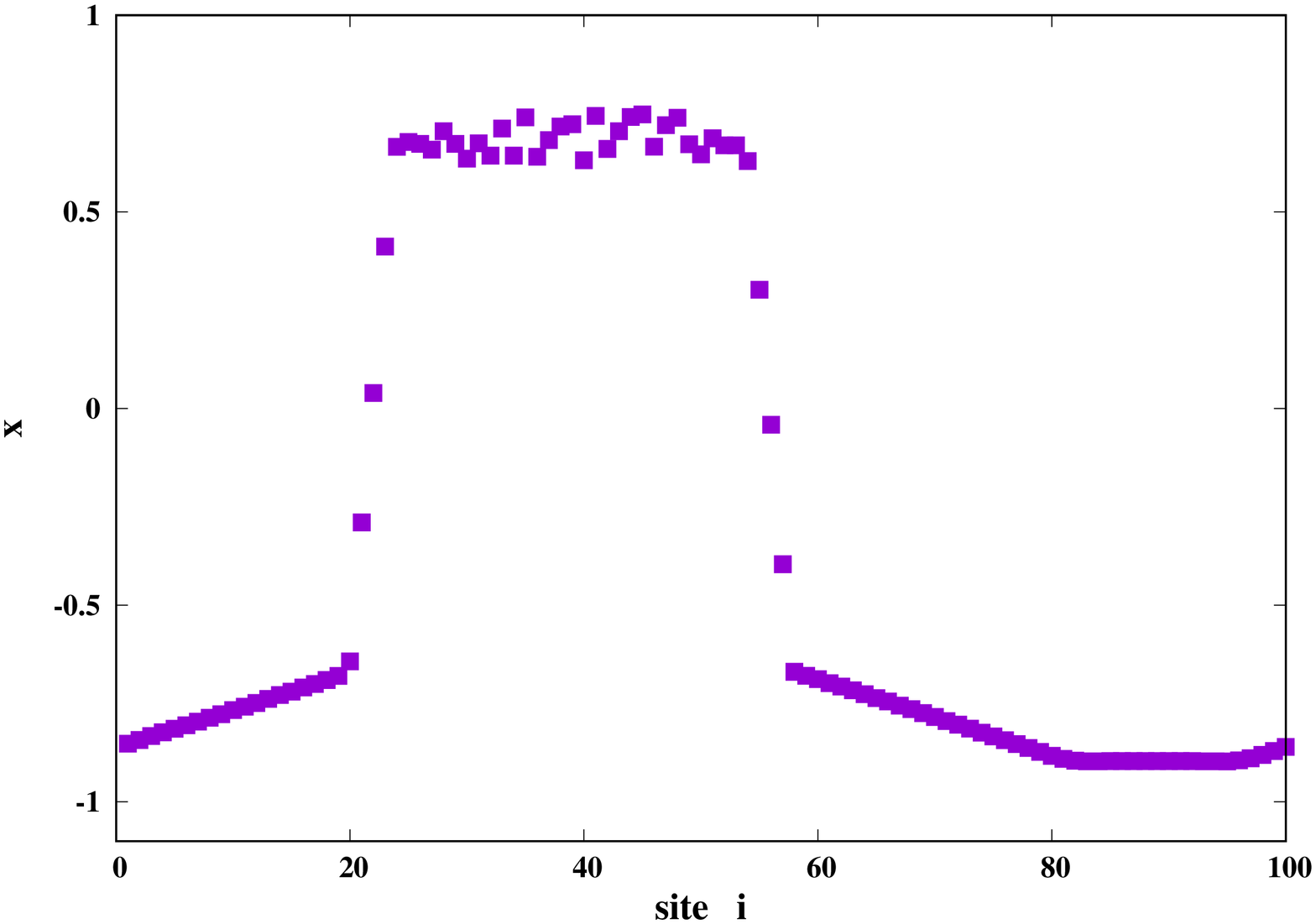} 
  \includegraphics[width=0.45\linewidth,height=0.75\linewidth,angle=270]{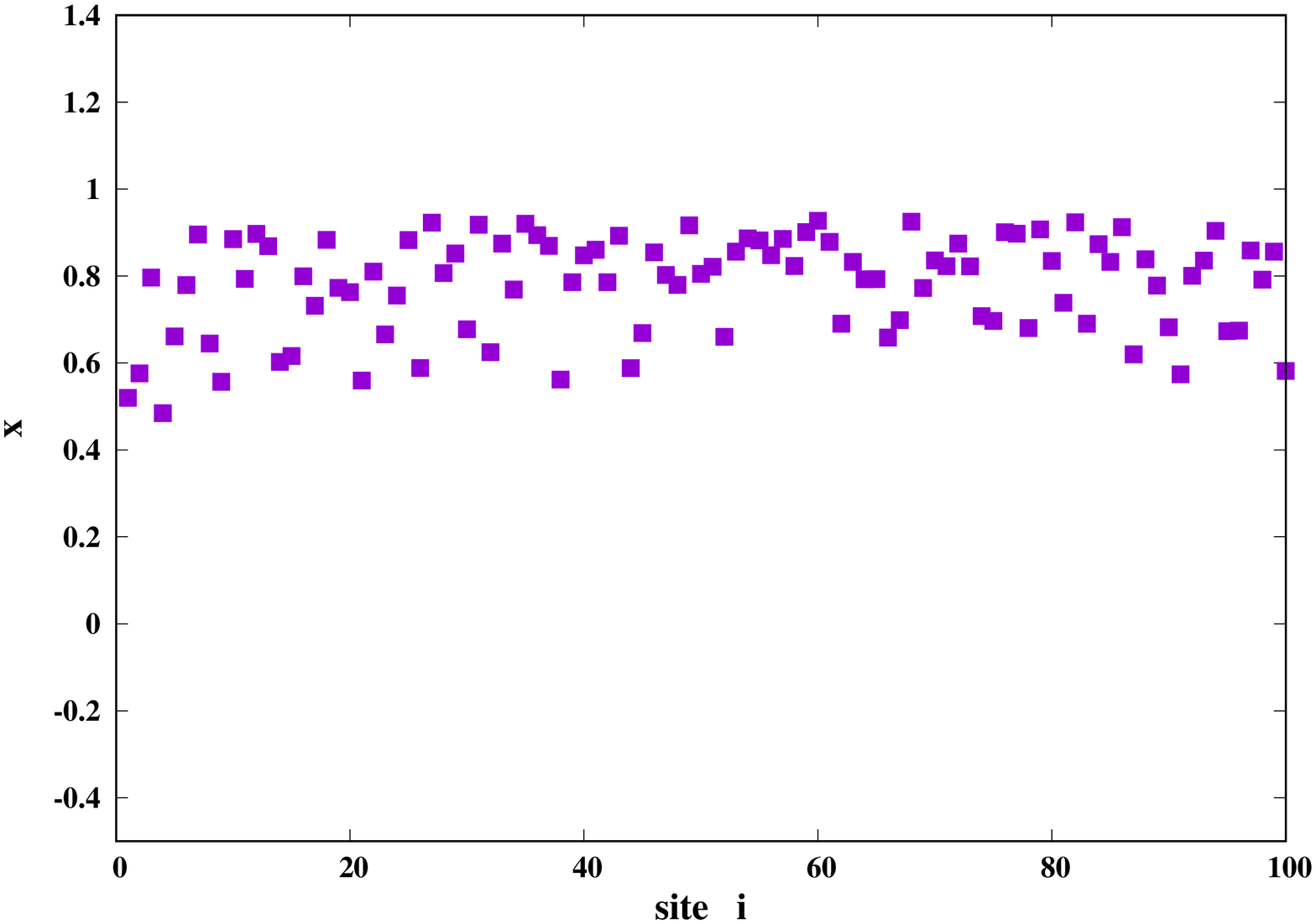} 
\caption{Spatial profile $x (i)$ of coupled piecewise-linear bistable maps ($i =1, \dots 100$), after 500 transient time steps, for fraction of random links: (top panel) $p=0$, and (lower panel) $p=0.3$. Here the parameters in Eqn.~\ref{pw_cml} are $p_1 = -0.5$, $p_2 = -2.4$, $l = 1.5$, $k=25$  and $\epsilon=0.35$. The initial state is exactly the same in both panels.\\}
\label{pw_0.35_p_0_0.3_x_i}
\end{figure}

We present yet another representative example of the destruction of a chimera state by random links in the ring of piecewise-linear maps in Fig.~\ref{pw_eps_0.35}. It is clearly evident that the presence of very few random links in the ring destroys the chimera-like pattern observed in the ring (see top panel of Fig.~\ref{pw_eps_0.35} vis-a-vis the middle panel), yielding a spatiotemporal fixed point. Increasing the fraction of random links here continues to be eliminate chimeras, yielding spatiotemporal chaos in a narrow band of state space. This can be seen more directly in the contrasting spatial profiles obtained for the case of $p=0$ and $p=0.3$ displayed in Fig.~\ref{pw_0.35_p_0_0.3_x_i}.

\begin{figure}[htb]  
\centering
\includegraphics[width=0.45\linewidth,height=0.75\linewidth,angle=270]{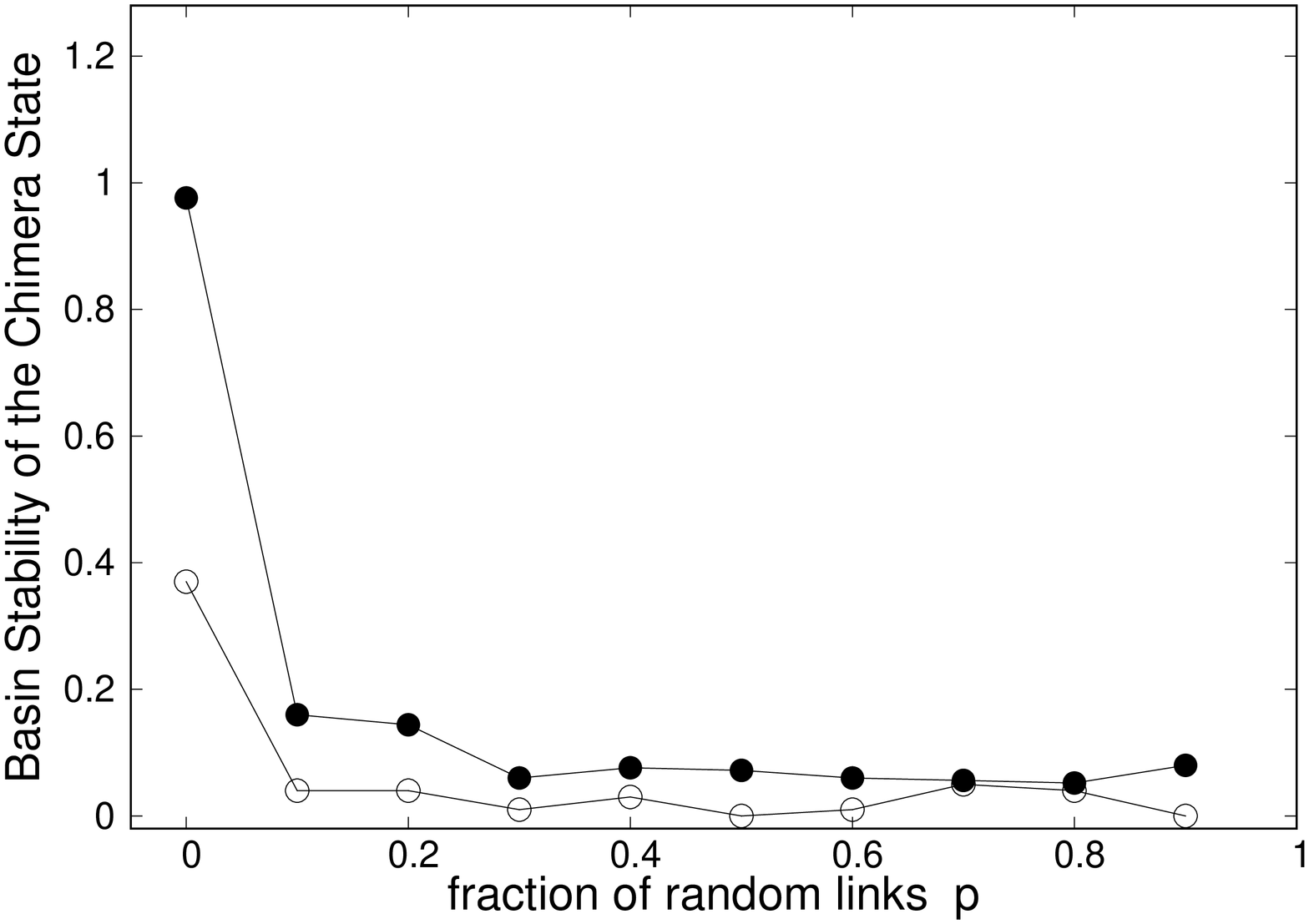}
\caption{Basin Stability of the chimera state in coupled piecewise-linear bistable maps, where Basin Stability is estimated from the fraction of initial states that go to the chimera state in the system of coupled piecewise linear maps. Here $\epsilon = 0.2$ (solid circles) and $0.25$ (open circles), and Basin Stability is obtained by sampling $250$ random initial conditions. State variables are considered synchronized if they are similar within an accuracy of $10^{-4}$, after transience of $10^3$. Similar results are obtained for the parameter sets $\epsilon = 0.3, 0.35$.\\}
\label{pw_bs}
\end{figure}

In order to quantify the global impact of random links we again estimate the Basin Stability of the chimera state, which reflects the
probability of a generic initial state evolving to a chimera state. Fig.~\ref{pw_bs} shows this fraction as a function of the
fraction of random links $p$ in the ring. It is clearly seen that the basin size of the chimera state sharply decreases to zero as the fraction of random links becomes non-zero. In contrast to the case of coupled circle maps, the chimera state is not a global attractor for the coupled piecewise-linear maps, and Basin Stability of the chimera state for a ring of such maps is less than $1$. However, it is still very clearly evident that on randomizing even a few links the Basin Size of the chimera state shrinks even further and rapidly becomes close to zero with increasing fraction of random links. Similar trends were observed for the other parameter sets investigated. So the general trend of random links destroying chimera states holds here as well, and there is a drastic reduction of the fraction of initial states yielding chimeras in the presence of a small number of random links.

So we find evidence of the rapid destruction of the basin of chimera states for $p \rightarrow 0$. This again indicates the huge effect of random links on chimera patterns in coupled nonlinear systems. This finding significantly impacts the potential observability of chimera states, as in naturally occuring scenarios a small number of links may get randomized from time to time.\\

\medskip

\noindent
{\bf Coupled Limit Cycle Oscillators:}\\
    
Lastly in order to further explore the generality of the observations above, we investigate another broad class of systems, namely a collection of coupled oscillators described by coupled nonlinear ordinary differential equations. Further in this class of systems, we consider yet another form of coupling: {\em conjugate coupling}.

\begin{figure}[htb]
  \centering
    \includegraphics[width=0.45\linewidth,height=0.75\linewidth,angle=270]{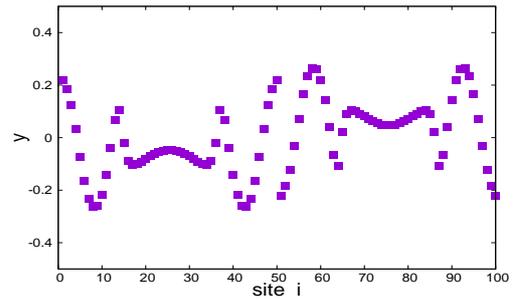}
  \includegraphics[width=0.45\linewidth,height=0.8\linewidth,angle=270]{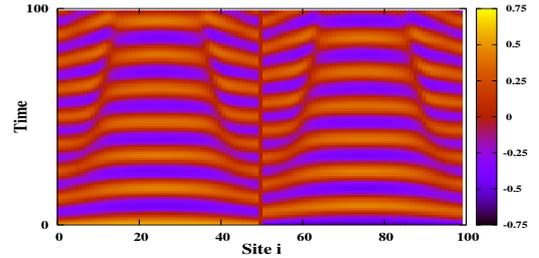} 
  \includegraphics[width=0.475\linewidth,height=0.8\linewidth,angle=270]{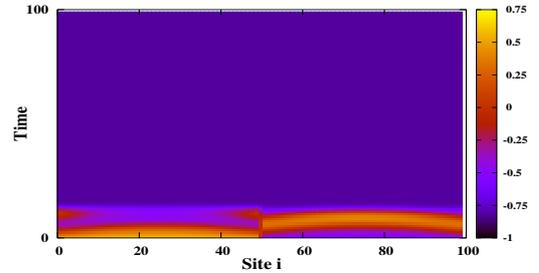}
\caption{Chimera-like spatial profile $x (i)$ ($i =1, \dots 100$) of conjugately-coupled Stuart-Landau oscillators, after $1000$ transient time steps, for a regular ring with $p=0$ (top panel); evolution of $x_i (t)$ of the oscillators, for $p=0$ (middle panel), and $p=0.01$, where at each instant, on an average, there is a {\em single} random link in the ring (lower panel). The system evolves from the exact same initial state in all panels. Parameters in Eqn.~\ref{SL} are $\omega = 0.5$, $\epsilon=0.9$ and $k = 35$.\\}
\label{SL_k_35_p_0}
\end{figure}

Specifically we consider a collection of prototypical Stuart-Landau (SL) oscillators. The Stuart-Landau oscillator is of broad relevance, as sufficiently close to any Hopf bifurcation, the variables with slower time-scales can be eliminated, yielding first order ordinary differential equations of the Stuart-Landau form. In our representative example we have conjugately coupled SL oscillators, that are nonlocally connected to $k$ neighbours on both sides, i.e. with range of coupling equal to $2 k$. So the dynamics of this system is given by $2N$ coupled nonlinear ordinary differential equations:
\begin{eqnarray}
\label{SL}
{\dot{x}}_{i} &= & (1 - x_i^2 - y_i^2) x_i - \omega y_i + \frac{\varepsilon}{2k} \sum_{j=i-k}^{i+k} \ [ y_j - x_i ] \\ \nonumber 
{\dot{y}}_{i} &= & (1 - x_i^2 - y_i^2) y_i + \omega x_i + \frac{\varepsilon}{2k} \sum_{j=i-k}^{i+k} \ [ x_j - y_i ] 
\end{eqnarray}
Here index $i$ specifies the site in the ring, with the local on-site dynamics being a Stuart-landau limit-cycle oscillator \cite{sl_chimera}.

\begin{figure}[htb]
    \centering
  \includegraphics[width=0.45\linewidth,height=0.75\linewidth,angle=270]{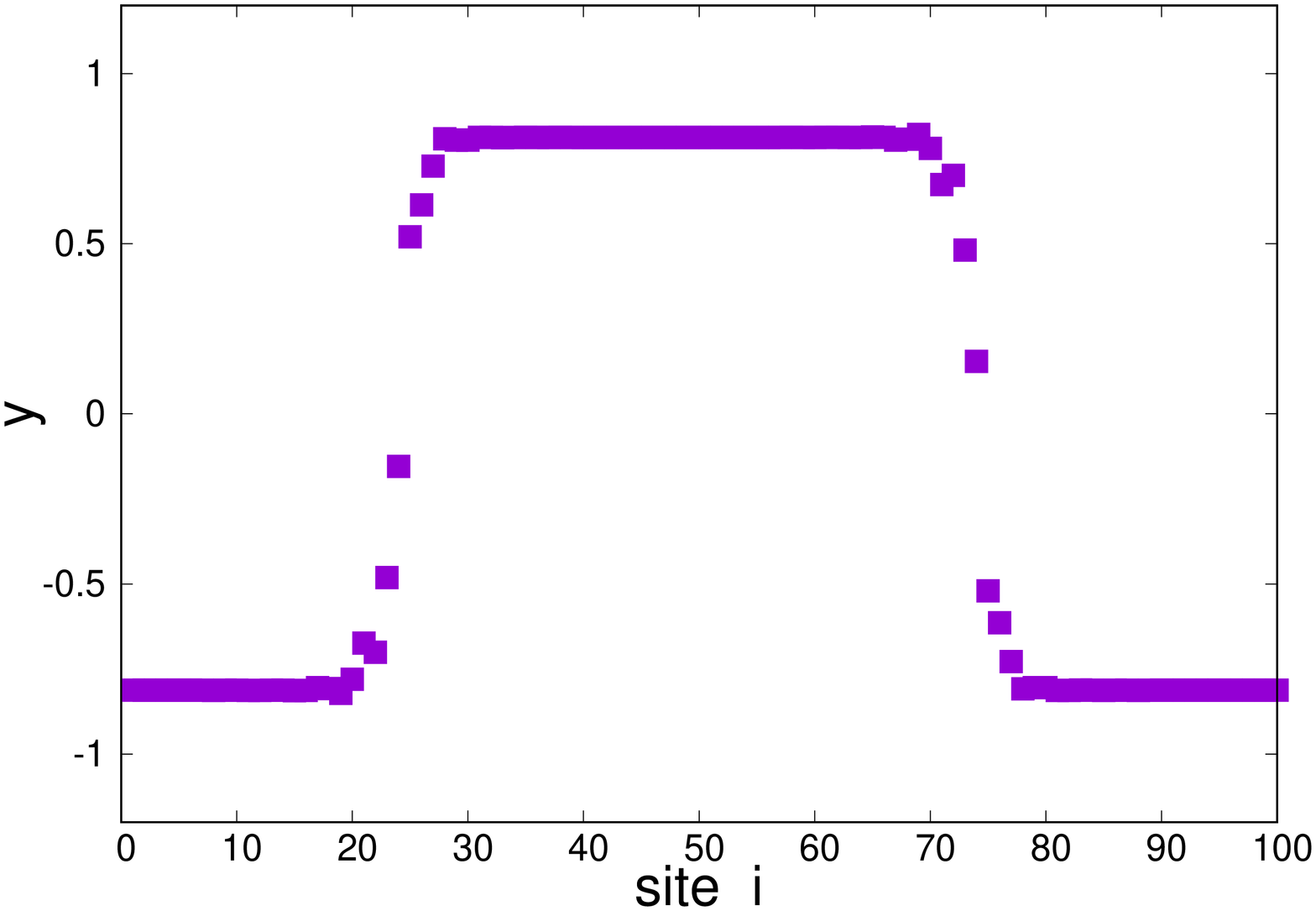}
  \includegraphics[width=0.45\linewidth,height=0.85\linewidth,angle=270]{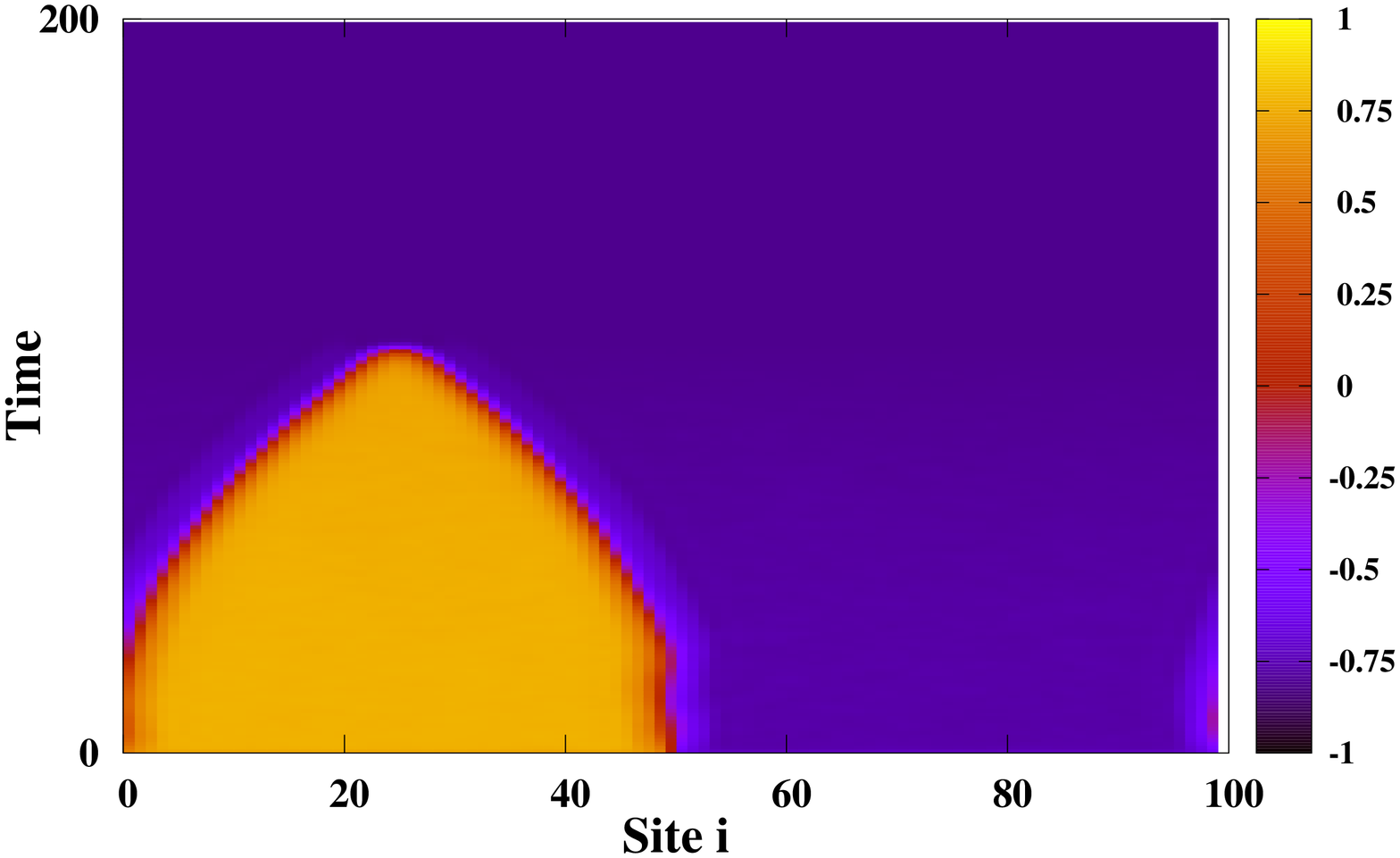}
  \caption{(Top panel) Spatial profile $x_i$ ($i =1, \dots 100$) for a regular ring of conjugately-coupled Stuart-Landau oscillators, (i.e. $p=0$) after $1000$ transient time steps; (lower panel) evolution of the system, with fraction of random links $p=0.1$, from the same initial state as in the top panel. Parameters in Eqn.~\ref{SL} are $\omega = 0.5$, $\epsilon=0.9$ and $k = 2$.\\}
\label{SL_k_2_p_0.1}
\end{figure}

\begin{figure}[htb]
      \centering
      \includegraphics[width=0.45\linewidth,height=0.75\linewidth,angle=270]{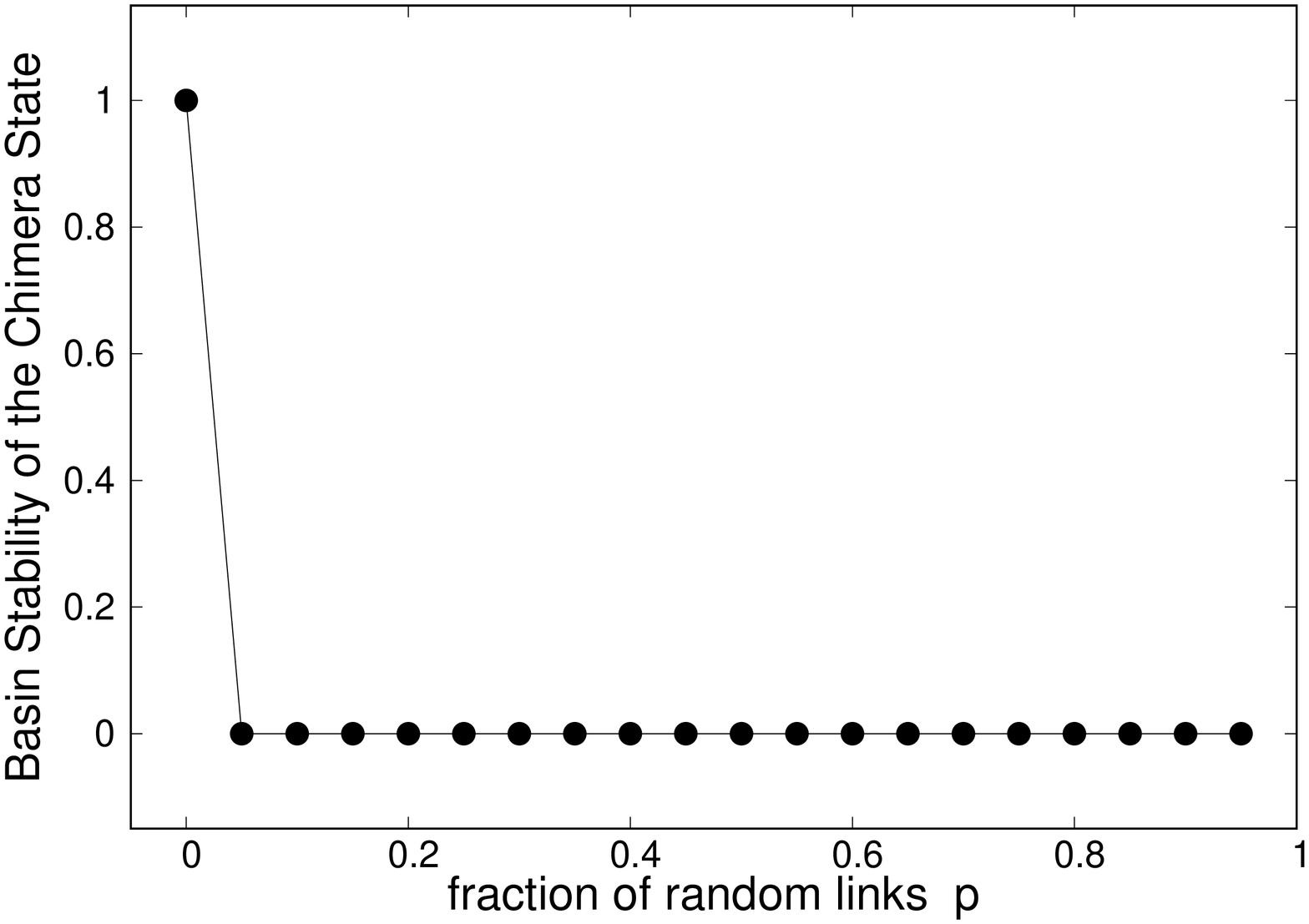}
      \caption{Basin Stability of the chimera state in the conjugately-coupled Stuart-landau oscillator system, where Basin Stability is estimated from the fraction of initial states that go to the chimera state, obtained by sampling $250$ random initial conditions. State variables are considered synchronized if they are similar within an accuracy of $10^{-6}$, after transience of $10^3$. Here parameters in Eqn.~\ref{SL} are $\omega = 0.5$, $\epsilon=0.9$ and $k = 35$. Identical results are obtained for the parameter set $\omega = 0.5$, $k=2$.\\}
\label{SL_bs}
 \end{figure}

In this example we investigate a {\em specific class of initial states}: half the ring ($i=1, \dots \frac{N}{2}$) has state ($x_0, y_0$) and the other half ($i= \frac{N}{2}+1, \dots N$) has state ($-x_0, -y_0$). We study parameter sets in Eqn.~\ref{SL} where all such initial states evolve to chimera states, as displayed in Fig.~\ref{SL_k_35_p_0} in a regular ring (i.e. for fraction of random links $p=0$). The lower panel of Fig.~\ref{SL_k_35_p_0} shows the evolution of the same collection of conjugately-coupled SL oscillator, with a {\em single} link randomized from time to time.  The initial state is the same as in the top panels of Fig.~\ref{SL_k_35_p_0}. Very clearly, the dynamical outcome is now drastically different, with the presence of the single random link in the ring destroying the chimera-like pattern observed in the ring, and yielding a spatiotemporal fixed point instead.

We present another representative example of the destruction of a chimera state by a single random link in Fig.~\ref{SL_k_2_p_0.1}. Here the oscillators are coupled to two nearest neighbours on both sides, i.e. $k=2$. Again it is clearly evident that the presence of very few random links in the ring destroys the chimera-like pattern (see top panel of Fig.~\ref{SL_k_2_p_0.1}), yielding a spatiotemporal fixed point instead (see lower panel of Fig.~\ref{SL_k_2_p_0.1}).

Again, to quantify the global impact of random links we estimate the Basin Stability of the chimera state, which reflects the probability of a generic initial state evolving to a chimera state. Fig.~\ref{SL_bs} shows this fraction as a function of the fraction of random links $p$ in the ring. It is clearly seen that the basin size of the chimera state sharply decreases sharply to zero as the fraction of random links becomes non-zero.\\

\noindent
{\bf Conclusions:}\\
    
In summary, we have studied the dynamics of a collection of coupled nonlinear systems, ranging from nonlinear maps to differential equations supporting limit cycles, under different coupling classes, connectivity ranges and initial states. Our focus in this work has been on the robustness of chimera states in the presence of a few time-varying random links. We find that the chimera states are often destroyed, yielding either spatiotemporal fixed points or narrow-band spatiotemporal chaos, in the presence of even a single dynamically changing random link. We also study the global impact of random links by exploring the Basin Stability of the  chimera state, and find that the basin size of the chimera state rapidly decreases to zero as the fraction of random links becomes finite, i.e. the transition to non-chimera states occurs in the limit of the fraction of random links tending to zero and very minimal spatial randomness is required to eliminate the chimera state. This indicates the far-reaching effect of a few switched links on chimera patterns in many systems, and impacts the potential observability of chimera states in naturally occurring scenarios.\\

\end{document}